# Cryogenic optical lattice clocks with a relative frequency difference of 1x10<sup>-18</sup>

Ichiro Ushijima[1,2,3*], Masao Takamoto[1,2,4*], Manoj Das[1,2,4], Takuya Ohkubo[1,2,3] & Hidetoshi Katori[1,2,3,4]

Time and frequency are the most accurately measurable quantities, providing foundations for science and modern technologies. The accuracy relies on the SI (Système International) second that refers to Cs microwave clocks with fractional uncertainties at $10^{-16}$ (Ref. 1). Recent revolutionary progress of optical clocks[1-5] aims to achieve $1 \times 10^{-18}$ uncertainty, which however has been hindered by long averaging-times[4,5] or by systematic uncertainties[1-3]. Here, we demonstrate optical lattice clocks with $^{87}$Sr atoms interrogated in a cryogenic environment to address the blackbody radiation-induced frequency-shift[6], which remains the primary source of clocks' uncertainties[1-3,7] and has initiated vigorous theoretical[8] and experimental[9,10] investigations. The quantum-limited stability[11] for $N{\sim}1,000$ atoms[2,3,12] allows investigation of the uncertainties at $2 \times 10^{-18}$ in two hours of clock operation. After 11 measurements performed over a month, the two cryo-clocks agree to within $(-1.1 \pm 1.6) \times 10^{-18}$. Besides its contribution to fundamental science[13] and the redefinition of the SI second[1,14], such a speedy comparison of accurate clocks provides a means for relativistic geodesy[15].

[1]Quantum Metrology Laboratory, RIKEN, Wako, Saitama 351-0198, Japan.
[2]Innovative Space-Time Project, ERATO, Japan Science and Technology Agency, Bunkyo-ku, Tokyo 113-8656, Japan.
[3]Department of Applied Physics, Graduate School of Engineering, The University of Tokyo, Bunkyo-ku, Tokyo 113-8656, Japan.
[4]RIKEN Center for Advanced Photonics, Wako, Saitama 351-0198, Japan.
*These authors contributed equally to this work.

The accuracy of an atomic clock relies on the presumed constancy of fundamental constants[16] and the decoupling of the electronic states from the ambient electromagnetic perturbations[17]. While clocks based on single ions, trapped near the zero of an electric quadrupole field, are expected to deliver the highest accuracy to date[4,5], the quantum projection noise (QPN)[11] requires days of averaging time to achieve their anticipated accuracy. In contrast, optical lattice clocks exploit well-engineered electromagnetic perturbation by the magic wavelength protocol[7,18] to facilitate the observation of many atoms $N_a$ simultaneously, thus reducing the averaging time $\tau$ by a factor of $N_a$. Clock stabilities approaching $10^{-16}(\tau/s)^{-1/2}$ have been demonstrated by rejecting the laser noise[12] or by applying lasers stabilized to long[2,3] or cryogenic[19] cavities. Such stabilities enable accessing $10^{-18}$ uncertainty in a few hours of clock operation, allowing extensive studies on systematic uncertainties, such as collisional shifts between spin-polarized fermions[20], and light shifts due to hyper-polarizability effects[21] and multipolar interactions of atoms with optical lattices[21,22]. Although the magic wavelength protocol allows cancellation of the lattice related Stark shift perturbations, the Stark shift caused by the ambient electric field[23], including the blackbody radiation (BBR), remains as a major perturbation in optical lattice clocks with Sr and Yb.

The BBR shift has long been recognized as a formidable adversary in pursuing high precision atomic clocks[6]. The difference of polarizabilities $\Delta\alpha = \alpha_e - \alpha_g$ between the excited ($\alpha_e$) and the ground ($\alpha_g$) states of the clock transition gives rise to the Stark shift $h\nu_{BBR} \approx -\frac{1}{2}\Delta\alpha \langle E^2 \rangle_T$ for the thermal field $\sqrt{\langle E^2 \rangle_T} \approx 8.3 \text{ V cm}^{-1}$ at $T = 300$ K. Among the operational optical clocks, a single Al$^+$ clock has the least sensitivity[4] to the thermal field, whereas neutral Yb and Sr clocks suffer from 300-700 times larger susceptibility[8] than Al$^+$ clocks. A Hg-based optical lattice clock, requiring deep-ultraviolet laser sources, has been proposed[24] and demonstrated[25] because of the reduced BBR shift. Generally, atoms easily accessible by visible or longer wavelength lasers tend to show higher susceptibility to the BBR owing to their relatively large static polarizabilities $\alpha_g$ and $\alpha_e$, unless they are cancelled out coincidentally. By utilizing a quantum logic scheme[16], optical clocks with highly charged ions[26], which have no dipole-allowed optical transitions, are proposed to drastically reduce the BBR perturbations.

Recently the BBR shift for the $^1S_0 - {}^3P_0$ clock transition of Sr was precisely evaluated to be $-2.277,8(23)$ Hz at $T = 300$ K based on a measured static polarizability difference $\Delta\alpha$ and a model of the dynamic contribution[9]. While the result allows correcting the room-temperature BBR shift with an uncertainty of $5 \times 10^{-18}$, reaching such an uncertainty requires evaluating the ambient temperature with an uncertainty smaller than $\Delta T = 70$ mK, which remains an experimental challenge. As the BBR energy density $\langle E^2 \rangle_T$ varies as $T^4$ (Stefan-Boltzmann law)[6], the BBR shift and its temperature dependence $(d\nu_{BBR}/dT \propto T^3)$ rapidly decreases with the surrounding temperature $T$. A cryogenic clock has been assumed in Ref. 7. For example, the BBR shift reduces to $\nu_{BBR} = -22$ mHz for $T = 95$ K, such that the relevant uncertainty of clocks can reach $1 \times 10^{-18}$ by determining the temperature to only within $\Delta T = 0.5$ K. Cryo-clocks therefore would make the BBR shift no longer a major obstacle in achieving $10^{-18}$ relative uncertainty. Similar strategy will be applicable to Yb clocks[10] as well.

Figure 1a shows our experimental setup. The cryogenic clock interrogates atoms inside a chamber which has a volume of $\sim 6$ cm$^3$ and can be cooled down to $95$ K by a Stirling refrigerator. The cryo-chamber is temperature-controlled by referencing a platinum resistance thermometer (see Methods). The cryo-chamber has two apertures with diameters of $\phi_1 = 0.5$ mm and $\phi_2 = 1$ mm to introduce atoms and lasers. The chamber is made of copper, with its inner surface black-coated to suppress multiple reflections of the room-temperature radiation that leaks through the apertures. The thickness of the coating is about 10-30 μm with sheet resistivity of less than $2 \times 10^6 \, \Omega \, \text{sq.}^{-1}$, which avoids accumulation of stray charges on the inner surface. Therefore the cryo-chamber also works as a Faraday cage to protect atoms from external electric fields[23].

Ultracold $^{87}$Sr atoms from a magneto-optical trap (MOT)[27] are loaded into a one-dimensional (1D) optical lattice with potential depth of 7 μK. The lattice consists of counter-propagating lasers tuned to the magic wavelength $\lambda_L \approx 813.4$ nm with their $e^{-2}$ waist diameter of 100 μm located in the middle of the cryo-chamber (see Fig. 1a). To adiabatically transport atoms over 23 mm from the MOT region to the cryo-chamber, one of the lattice lasers is frequency chirped for 60 ms (see Fig. 1c). We apply maximum frequency chirp of $\delta = 2$ MHz, which corresponds to a peak velocity of 0.8 m s$^{-1}$.

After transporting the atoms inside the chamber, the atoms are spin-polarized in the $^3P_0(F = 9/2, m_F)$ metastable state with $m_F$ either $9/2$ or $-9/2$ (see Fig. 1b and Methods). For the clock interrogation, the $^3P_0(F = 9/2, m_F) \rightarrow {}^1S_0(F = 9/2, m_F)$ transition is excited by a clock laser at $\lambda_C = 698$ nm. The clock laser is frequency stabilized to a 40-cm-long reference cavity made up of ultralow-expansion glass spacer and fused silica mirrors, and has an expected stability of $\sim 1 \times 10^{-16}$ at 1 s. After the interrogation period, the moving lattice transports the atoms back to the MOT region, where the population in the $^1S_0$ state is measured by driving the $^1S_0 - {}^1P_1$ transition at 461 nm.

Two clock setups are developed, namely Sr-1 and Sr-2, with (cryo-)chambers temperature-controlled at $T_1$ and $T_2$, respectively. Figure 2a shows the clock excitation spectra obtained by simultaneously scanning the clock laser frequencies of Sr-1 maintained at $T_1 = 95$ K (blue circles) and Sr-2 at $T_2 = 296$ K (orange circles) with a 400-ms-long π-pulse excitation, which resolves the room-temperature BBR shift of about 2 Hz.

To stabilize the clock laser frequency to the respective resonances $\nu_j(T_j)$ of Sr-$j$ operated at temperature $T_j$ with $j = 1$ or 2, we measure the excitation probability difference $\Delta p = p_+ - p_-$ of the Rabi profile at detunings of $\Delta = \pm(\gamma/2)$ with $\gamma \approx 2.7$ Hz for a 300-ms-long π-pulse (Fig. 1c). In addition, we implement Zeeman and vector light shift cancellation[28] (see Methods), which requires 4 measurement cycles (each with a cycle time of $T_C \approx 1.5$ s) to obtain the clock frequency $\nu_j(T_j)$. The inset of Fig. 2a shows the BBR induced frequency difference $\Delta\nu(T_1, T_2) = \nu_2(T_2) - \nu_1(T_1)$ between clocks operating at temperatures $T_1$ and $T_2$. The blue line is measured for $\Delta\nu(95\text{ K}, 95\text{ K})$ with both clocks at cryogenic temperatures, while the orange line shows $\Delta\nu(95\text{ K}, 296.2\text{ K})$. By averaging the data for $\approx 3 \times 10^4$ s, we determined the differential BBR shift to be $\Delta\nu(95\text{ K}, 296.2\text{ K}) = -2.138{,}4(21)$ Hz.

Figure 2b shows the temperature dependence of $\Delta\nu(95\text{ K}, T_2)$ by varying the temperature of Sr-2 from $T_2 = 95$ K to $296.2$ K (black circles), while Sr-1 operates at $T_1 = 95$ K as a reference. The temperature dependence of the BBR shift is given by

$$\nu_{\text{BBR}}(T) = \nu_{\text{stat}}(T/T_0)^4 + \nu_{\text{dyn}}(T/T_0)^6 + \mathcal{O}(T/T_0)^8,$$

where $\nu_{\text{stat}}$ and $\nu_{\text{dyn}}$ are the static and dynamic contributions[8,9] at $T_0 = 300$ K. The data points follow the temperature dependence of $\Delta\nu(95\text{ K}, T) = \nu_{\text{BBR}}(T) - \nu_{\text{BBR}}(95\text{ K})$. Using $\nu_{\text{stat}}^{\text{PTB}} = -2.130{,}23$ Hz as the static contribution, obtained from the measurements carried out at the Physikalisch-Technische Bundesanstalt[9], the best fit is obtained for $\nu_{\text{dyn}} = -0.148{,}0(26)$ Hz, as indicated by the red line. The upper panel of Fig. 2b shows the residuals in the fitting, where shaded region shows the $1\sigma$ error for the fit for $\nu_{\text{dyn}}$.

To evaluate cryo-clocks' uncertainties, we synchronously[12] operate the two clocks at $T_1 = T_2 = 95$ K. Figure 3a shows the Allan standard deviation for $\Delta\nu(95\text{ K}, 95\text{ K})$, which falls as $\sigma_y(\tau) = 1.8 \times 10^{-16} \, (\tau/\text{s})^{-1/2}$ (blue dashed line) and reaches $2.3 \times 10^{-18}$ for an averaging time $\tau = 6{,}000$ s. This stability is close to the QPN limited stability (red dashed line) for $N_a = 1{,}000$ atoms that is estimated from laser-induced fluorescence measurements. Figure 3b shows the frequency difference between Sr-1 and Sr-2 with the averaging time for each data point lying between $\tau = 4{,}000 - 10{,}000$ s. After averaging 11 separate measurements performed over a month, we obtained a frequency difference of $\Delta\nu = -0.5 \pm 0.7$ mHz or $\Delta\nu/\nu_0 = (-1.1 \pm 1.6) \times 10^{-18}$ with $\nu_0 \approx 429$ THz the clock transition frequency. The red dashed line and the shaded region show the averaged frequency and its $1\sigma$ uncertainty, respectively. Table 1 gives the uncertainty budget. The total systematic correction for each clock is evaluated to be 73.9 mHz. A detailed description is given in Methods.

In summary, we have developed a pair of cryogenic Sr optical lattice clocks that demonstrate an agreement at a fractional frequency uncertainty of $1.6 \times 10^{-18}$. The cryo-clocks enable us to directly investigate the BBR shift of $\nu_{\text{BBR}}(300\text{ K}) = -2.278{,}2(26)$ Hz, which agrees well with the previous indirect measurements[9]. Synchronous comparison of two cryo-clocks allowed us to evaluate clock uncertainties at $2 \times 10^{-18}$ within an averaging time of 2 hours. Such stabilities are useful in comparing two distant clocks connected by a fibre link[29], which will open up new possibilities for measuring the gravitational clock shift[15,30] as a tool for geodesy.

**Methods**

**Uncertainties of BBR shift for a cryogenic $^{87}$Sr optical lattice clock.**

The temperature of the chamber is measured by platinum resistance thermometers, which are calibrated with uncertainties of 22 mK and 62 mK at $T = 77$ K and $T = 300$ K respectively. Note that while the calibration uncertainty provides a negligible contribution of $2.5 \times 10^{-20}$ for $T = 77$ K, it amounts to as much as $4.5 \times 10^{-18}$ for $T = 300$ K. For the Stirling pump to reach 95 K, we supply an alternating current (corresponding power is about 30 W) to a compressor, which is placed inside a μ-metal shielding box to suppress ac magnetic field (see photo in Fig. 1a). At $T = 95.00(4)$ K, the blackbody shift caused by the cryo-chamber wall is $\nu_{\text{BBR}}^{95\,\text{K}} = -21.55(4)$ mHz. Atoms in the cryo-chamber are exposed to the room temperature (RT) radiation that leaks through the two apertures (diameters of $\phi_1 = 0.5$ mm and $\phi_2 = 1$ mm). The corresponding BBR shift is $\nu_{\text{BBR}}^{\text{RT}} = -1.54(33)$ mHz by moving the atoms into the chamber by 10 mm, where we assume a room temperature of $T_{\text{RT}} = 296(5)$ K and the total solid angle of the apertures $\delta\Omega_{\text{RT}} = 9(2)$ msr. In addition, reflections of the penetrated RT radiation by the chamber inner wall cause BBR shift[31]. To reduce this contribution, the inner wall of the cryo-chamber is black-coated with hemispherical reflectance of $\approx 2$ % for $\lambda = 2 - 11$ μm and less than 10 % for $\lambda = 1 - 30$ μm. Assuming an emissivity of $\epsilon = 0.9(1)$, this contribution is as small as $\nu_{\text{BBR}}^{\text{RT(r)}} = -0.17(17)$ mHz. We therefore estimate the total BBR shift to be $\nu_{\text{BBR}}^{95\,\text{K}} + \nu_{\text{BBR}}^{\text{RT}} + \nu_{\text{BBR}}^{\text{RT(r)}} = -23.26(37)$ mHz.

**State preparation of spin-polarized $^{87}$Sr atoms inside the cryo-chamber.**

After transporting the atoms inside the cryogenic chamber, a π-polarized pumping laser at 689 nm, superimposed on the lattice laser, drives the $^1S_0$ ($F =$ $9/2, m_F) - {}^3P_1$ $(F = 7/2, m_F)$ transition to pump the atomic population into the two stretched states $m_F = \pm 9/2$ in the ${}^1S_0$ $(F = 9/2)$ state. Thereafter the atoms in either of the stretched states $m_F = \pm 9/2$ are selectively excited to the ${}^3P_0$ $(F = 9/2, m_F = \pm 9/2)$ metastable state by applying a 40-ms-long π-pulse resonant to the corresponding Zeeman transition. A bias magnetic field of $|\mathbf{B}| = 48$ µT is applied throughout the interrogation period to resolve the adjacent Zeeman transitions (~ 50 Hz) by the Rabi excitation linewidth of 20 Hz. The unexcited atoms in the ${}^1S_0$ state are then blown out of the lattice by exciting the ${}^1S_0 - {}^1P_1$ transition at 461 nm. Although half of the lattice-trapped atoms are lost in this spin-polarization scheme, this process can be implemented using lasers aligned collinear with the lattice, thereby removing any need for additional apertures causing BBR shift. This spin purification process[32] allows nearly 100 % spin-polarization in the ${}^3P_0$ state, enabling more than 90 % excitation (see Fig. 2a).

**Doppler shift cancellation system for clock spectroscopy.**

Relative motion of atoms confined in the optical lattice against the clock laser causes a Doppler shift[12], which degrades the stability as well as accuracy. To suppress the Doppler shift, a DCC (Doppler Cancellation for Clock laser) unit in Fig. 1a stabilizes the path length for the clock laser by monitoring a reflection from a facet of a beam splitter cube (BSC) that serves as a reference for the lattice standing wave. Sharing the same BSC, a DCL (Doppler Cancellation for Lattice lasers) unit stabilizes the relative phase of counter propagating lattice lasers using a frequency shifter FS4. During atom transport by a moving lattice, the feedback loop for DCL is blocked and a constant frequency is fed to FS4, while FS5 provides the necessary frequency offset to operate the moving lattice. Note that upper shaded region of the BSC does not share the same optical path for the clock and lattice lasers. The unshared optical path length $l_{op}$ ($\approx$ 3.8 cm) causes the Doppler shift of $\frac{\nu_D}{\nu_0} = \frac{1}{c}\frac{dl_{op}}{dt}$ with $c$ the speed of light. For example, a linear temperature drift of 1 K of BSC made of BK7 in 1,000 s corresponds to the Doppler shift of $\langle \frac{1}{c}\frac{dl_{op}}{dt}\rangle_{\tau=1000\,s} \approx 10^{-18}$, which gives negligible contribution to this measurement.

**Corrections and uncertainties for frequency comparison between two cryogenic clocks.**

The first order Zeeman shift and the vector light shift are cancelled out[28] after four clock cycles by calculating $\nu_0 = (\nu_z^+ + \nu_z^-)/2$, where the Zeeman components $\nu_z^{\pm}$ correspond to the $^1S_0$ ($F = 9/2, m_F = \pm 9/2$) − $^3P_0$ ($F = 9/2, m_F = \pm 9/2$) transition frequencies respectively (see Figs. 1b and 1c). The quadratic Zeeman shift then contributes the largest correction, which is calculated from the first order Zeeman shift $\Delta\nu_z = \nu_z^+ - \nu_z^-$. Table 1 lists the typical shift for a bias magnetic field of $|\mathbf{B}| =$ 48 µT, where the uncertainty arises from the coefficient for the quadratic Zeeman shift.

The uncertainty in lattice light shift is evaluated to be 1.7 mHz by measuring the clock frequency shift in time-interleaved sequence where we alternately reduce the lattice intensity to 50 %, which also determines the uncertainty in the magic wavelength. According to Ref. 21, the lattice light shift and relevant uncertainty arising from hyperpolarizability and multipolar effects are estimated to be $-0.7(1.6)$ mHz for a potential depth of $40\,E_R$ and a mean vibrational state occupation of $\langle n \rangle = 0.2$ with $E_R$ being the recoil energy of lattice laser. As this estimated uncertainty is within our measurement uncertainty, we represent the lattice light shift uncertainty based on our measurements. In a clock comparison, most of the lattice light shifts are cancelled out as the two clocks share the same lattice laser with an intensity difference of less than 15 %.

The imbalance of counter-propagating lattice laser intensities introduces the travelling wave component, which causes the multipolar (M1/E2) light shift[7,22] and affects the magic wavelength. The relevant uncertainty, which is referred to as the travelling wave contamination, is evaluated by the clock frequency shift when alternately increasing the intensity of one of the counter-propagating lattice lasers by 50 %.

The clock light shift is calculated for a 300-ms-long $\pi$ pulse with intensity uncertainty of 20 %.

The first order Doppler shift arises from the optical path of about 15 cm that are used to split the clock laser and to deliver to DCC units of Sr-1 and Sr-2.

The typical number of atoms of $N_{\text{typ}} \approx 1{,}000$ used for the present clock operation corresponds to a single-lattice-site occupancy of $\approx 0.8$ atoms. We measured atom-number-dependent clock-shifts by alternately reducing the number of atoms $N_{\text{typ}}$ down to $N_{\text{low}}(\approx N_{\text{typ}}/3)$ in Sr-1(2), while keeping the number in Sr-2(1) fixed at $N_{\text{typ}}$. The observed density shift is $0.4(1.8)$ mHz for $N_{\text{typ}}$, which is within the measurement uncertainty.

The height difference between the two clocks is within 1 mm and no appreciable gravitational shift appears in the budgets.

**Acknowledgements:** This work received support partly from the JSPS through its FIRST Program and from the Photon Frontier Network Program of MEXT, Japan. The authors thank N. Nemitz, T. Takano, A. Yamaguchi and N. Ohmae for useful comments and conversations.

**Author Contributions:** H.K. envisaged and initiated experiments. H.K., M.T., M.D., T. O. and I.U. designed apparatus and experiments. I.U. and M.T. carried out experiments and analysed data. I.U., M.T. and H.K. discussed the results and contributed to the writing of the draft.

**Author Information:** Correspondence and requests for material should be addressed to H. K. (e-mail: hkatori@riken.jp).

| Effect | Correction (mHz) | Uncertainty (mHz) Relative (Absolute) | |
|---|---|---|---|
| Quadratic Zeeman shift | 50.2 | 0.05 | (0.4) |
| Blackbody radiation shift | 23.3 | 0.6 | (0.4) |
| Lattice light shift | 0 | 0.3 | (1.7) |
| Travelling wave contamination | 0 | 1.4 | (1.0) |
| Clock light shift | 0.02 | 0.006 | (0.01) |
| First order Doppler shift | 0 | 0.3 | (0.2) |
| AOM chirp & switching | 0 | < 0.1 | (< 0.1) |
| Servo error | 0 | 0.2 | (1.5) |
| Density shift | 0.4 | 1.0 | (1.8) |
| Systematic total | 73.9 | 1.9 | (3.1) |

**Table 1 | Uncertainty budget**

'Relative' lists the residual contributions that are not cancelled out for the two clocks, owing to either intensity fluctuation of the lasers or to the fluctuations in the atom number. 'Absolute' estimates the uncertainty for each individual clock.

**Figures**

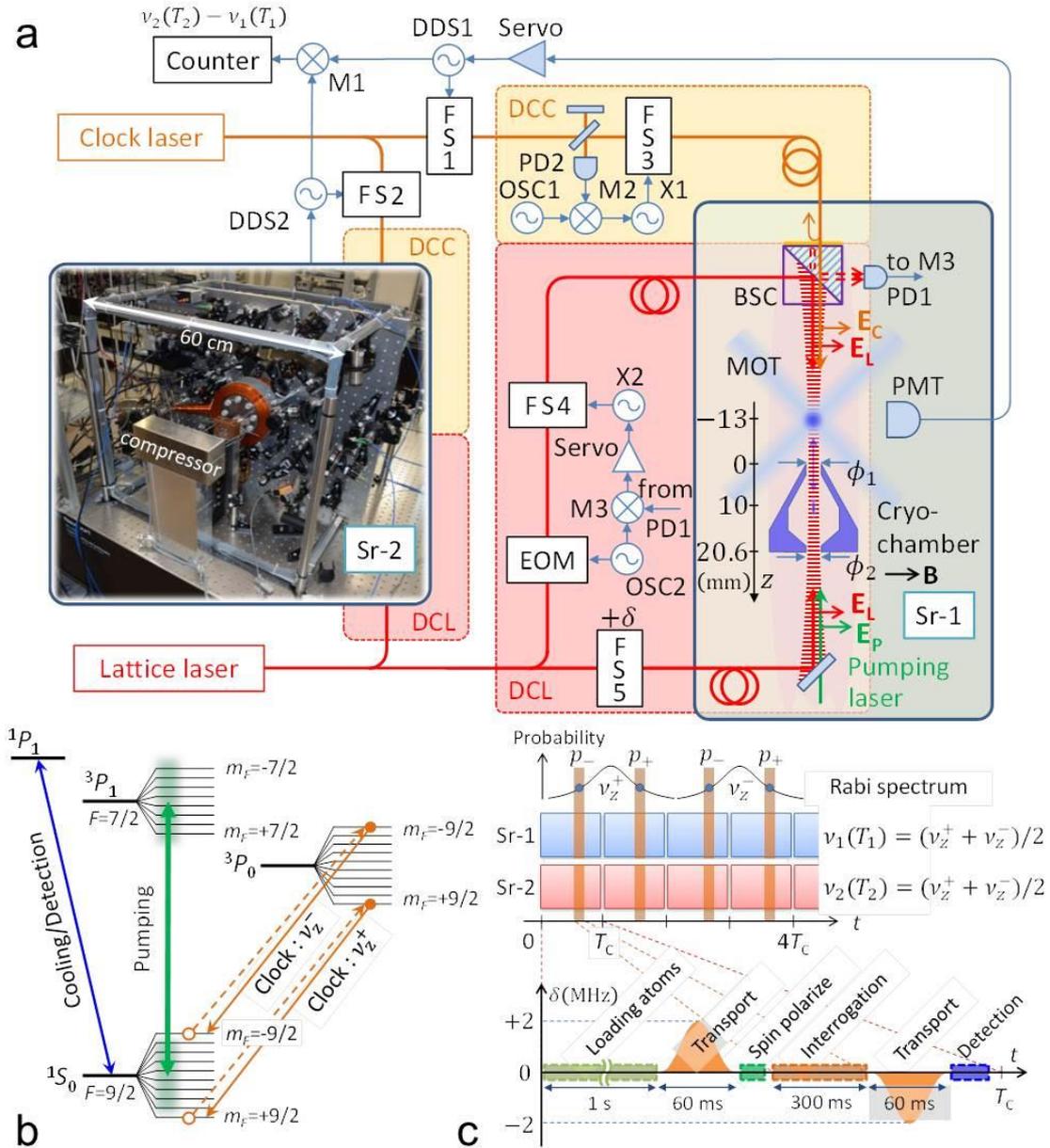

**Figure 1 | Experimental setup of cryogenic optical lattice clocks. a,** Two cryo-clocks, Sr-1 and Sr-2, operate by sharing the clock and lattice lasers delivered by optical fibres. Ultracold $^{87}$Sr atoms trapped in a magneto-optical trap (MOT) are transported by a moving lattice into a cryo-chamber and spin-polarized in the $^3P_0$ state. After interrogating the $^3P_0 \to {}^1S_0$ clock transition,
13

atoms are transported back to the MOT region to measure atomic population by laser-induced fluorescence with a photomultiplier tube (PMT). The deduced excitation probabilities for Sr-1 and Sr-2 are used to steer the respective clock-laser frequencies by frequency shifters (FS1,FS2) driven by direct digital synthesizers (DDS1,DDS2). By referencing a beam splitter cube (BSC), we implement a Doppler cancellation for the clock laser (DCC) and for the lattice lasers (DCL). DCC stabilizes the effective beam path length to the clock laser by FS3. DCL keeps the relative phase of the counter propagating lattice lasers using FS4 driven by a voltage-controlled crystal oscillator (X2), where an electro-optic modulator (EOM) is used to generate Pound-Drever-Hall error signal using a photo detector (PD1) demodulated by a mixer (M3). The same DCC and DCL are installed in Sr-2 as well, with minimized distance between the two DCCs $(\approx 15 \text{ cm})$ to reduce the relative motion of Sr-1 and Sr-2. Each clock setup is constructed inside a volume of $(60 \text{ cm})^3$ as shown in the photo for Sr-2, and is covered by black walls. **b,** Relevant energy levels for [87]Sr. **c,** Timing chart for the synchronous clock operation with cycle time of $T_C = 1.5 \text{ s}$. Using 4 cycles that probe four shoulders of the Rabi profile with an interrogation time of $T_i = 300 \text{ ms}$, we obtain clock frequency by rejecting Zeeman and vector light shift. The atoms are transported over $23 \text{ mm}$ by applying a frequency chirp of $\delta = 2 \text{ MHz}$ to FS5.

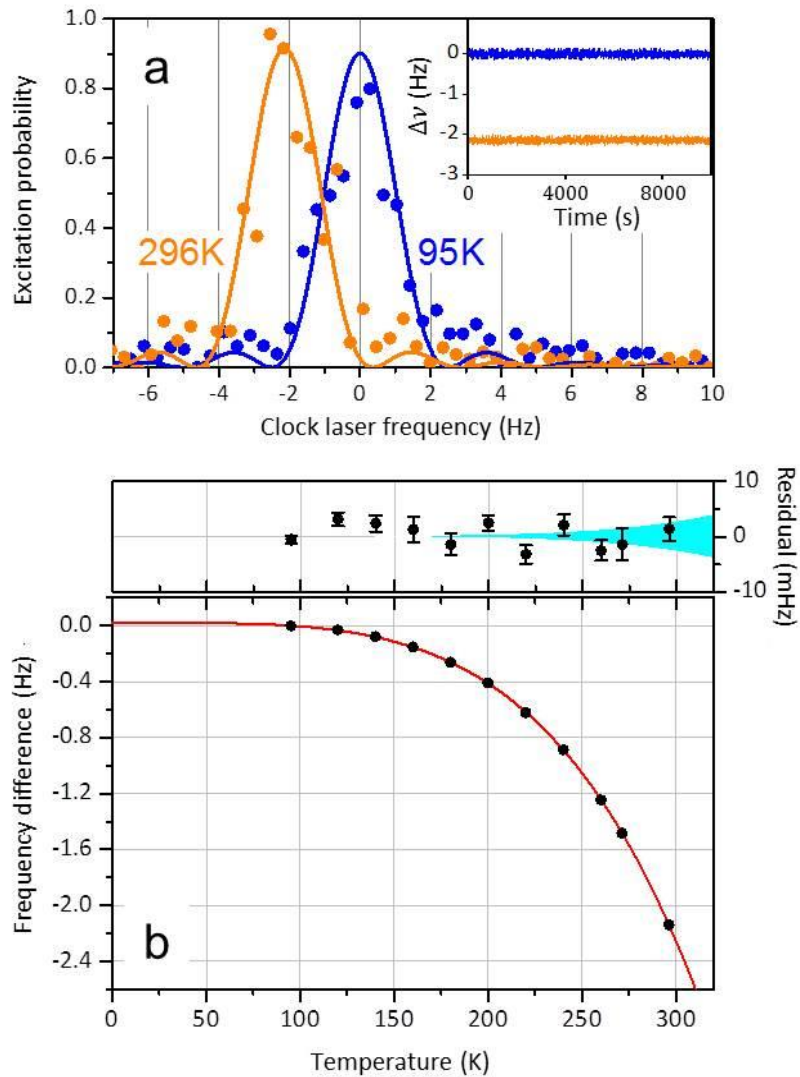

**Figure 2 | Temperature dependent BBR shift. a,** Clock excitation spectra measured for $T = 95$ K (blue circles) and $296$ K (orange circles). Solid lines show the Rabi spectra calculated for a 400-ms-long π-pulse. Inset shows the frequency difference $\Delta \nu(95\text{ K}, T_2)$ with $T_2 = 95$ K (blue) and $T_2 = 296.20$ K (orange), which determines the BBR shift of $\Delta \nu(95\text{ K}, 296.20\text{ K}) = -2.138{,}4(21)$ Hz. **b,** The frequency difference $\Delta \nu(95\text{ K}, T_2)$ between two clocks (black circles) measured by changing the temperature $T_2$ of Sr-2, while keeping the temperature of Sr-1 at $T_1 = 95$ K. Red line shows the fit to the data points to

obtain a dynamic contribution of $\nu_{\text{dyn}} = -0.148,0(26)$ Hz. Upper panel shows the residuals in the fitting. Error bars include statistical and temperature measurement uncertainties. Shaded region corresponds to $1\sigma$ uncertainties for $\nu_{\text{dyn}}$.

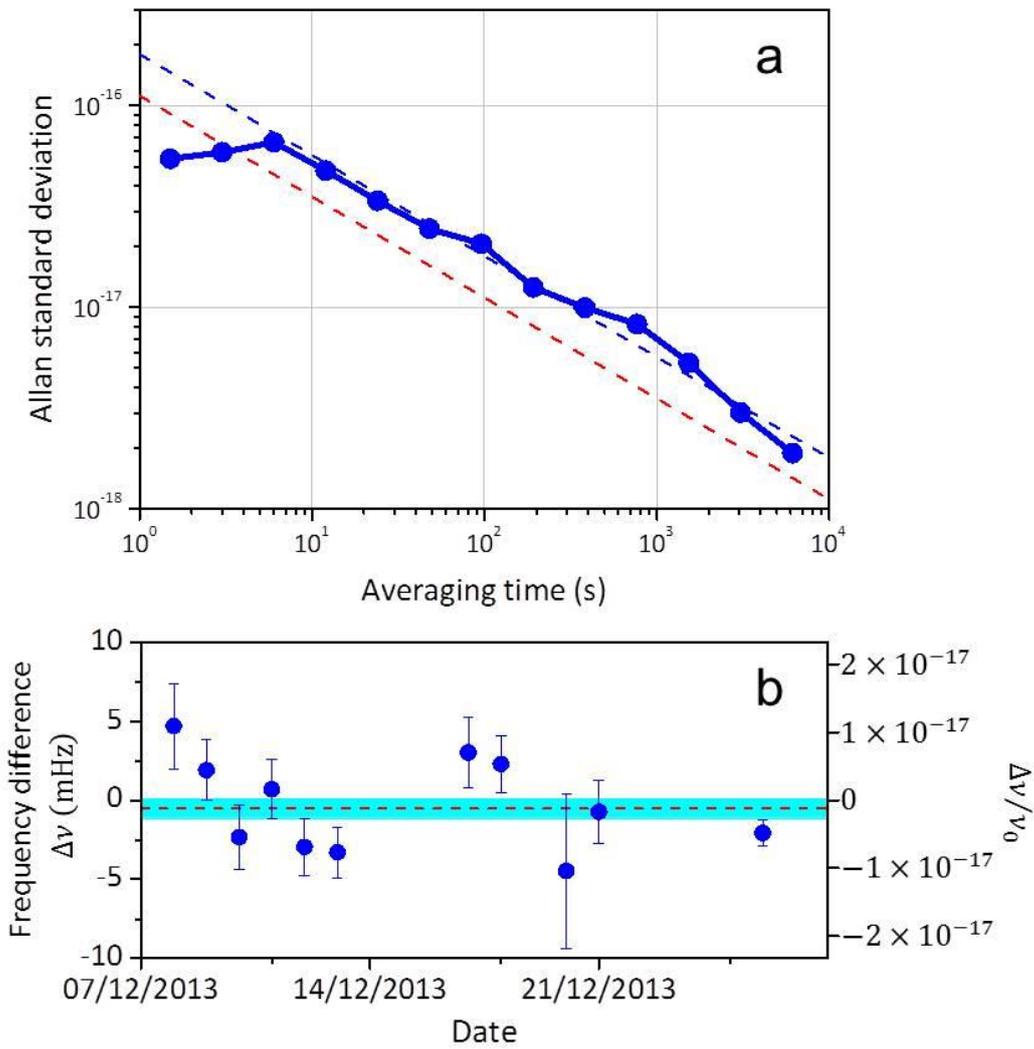

**Figure 3 | Frequency comparison between two cryo-clocks. a,** Allan standard deviation for the two cryo-clocks, falling as $\sigma_y(\tau) = 1.8 \times 10^{-16} \, (\tau/s)^{-1/2}$ (blue dashed line) and reaching $2 \times 10^{-18}$ for an averaging time $\tau = 6,000 \, s$. Red dashed line shows the calculated QPN limited stability for

$N_\mathrm{a} = 1{,}000$ atoms. **b,** The observed frequency differences between the two cryo-clocks. Error bars represent the $1\sigma$ statistical uncertainty. After averaging 11 measurements performed over a month, the two clocks agree to within $\Delta\nu = -0.5 \pm 0.7$ mHz, or $\Delta\nu/\nu_0 = (-1.1 \pm 1.6) \times 10^{-18}$ relative uncertainty with $\nu_0$ being the clock frequency. Red dashed line presents the average frequency and shaded region the $1\sigma$ statistical uncertainty.

References

1   Le Targat, R. *et al.* Experimental realization of an optical second with strontium lattice clocks. *Nat Commun* **4** (2013).

2   Hinkley, N. *et al.* An atomic clock with $10^{-18}$ instability. *Science* **341**, 1215-1218 (2013).

3   Bloom, B. J. *et al.* An optical lattice clock with accuracy and stability at the $10^{-18}$ level. *Nature* **506**, 71-75 (2014).

4   Chou, C. W., Hume, D. B., Koelemeij, J. C. J., Wineland, D. J. & Rosenband, T. Frequency Comparison of Two High-Accuracy Al$^+$ Optical Clocks. *Phys. Rev. Lett.* **104**, 070802 (2010).

5   Huntemann, N. *et al.* High-Accuracy Optical Clock Based on the Octupole Transition in $^{171}$Yb$^+$. *Phys. Rev. Lett.* **108**, 090801 (2012).

6   Itano, W. M., Lewis, L. L. & Wineland, D. J. Shift of $^2S_{1/2}$ hyperfine splittings due to blackbody radiation. *Phys. Rev. A* **25**, 1233-1235 (1982).

7   Katori, H., Takamoto, M., Pal'chikov, V. G. & Ovsiannikov, V. D. Ultrastable optical clock with neutral atoms in an engineered light shift trap. *Phys. Rev. Lett.* **91**, 173005 (2003).

8   Porsev, S. G. & Derevianko, A. Multipolar theory of blackbody radiation shift of atomic energy levels and its implications for optical lattice clocks. *Phys. Rev. A* **74**, 020502 (2006).

9   Middelmann, T., Falke, S., Lisdat, C. & Sterr, U. High Accuracy Correction of Blackbody Radiation Shift in an Optical Lattice Clock. *Phys. Rev. Lett.* **109**, 263004 (2012).

10  Sherman, J. A. *et al.* High-Accuracy Measurement of Atomic Polarizability in an Optical Lattice Clock. *Phys. Rev. Lett.* **108**, 153002 (2012).

11  Itano, W. M. *et al.* Quantum projection noise: Population fluctuations in two-level systems. *Phys. Rev. A* **47**, 3554-3570 (1993).

12  Takamoto, M., Takano, T. & Katori, H. Frequency comparison of optical lattice clocks beyond the Dick limit. *Nature Photon* **5**, 288-292 (2011).

13  Bordé, C. J. Base units of the SI, fundamental constants and modern quantum physics. *Phil. Trans. Roy. Soc. A* **363**, 2177-2201 (2005).

14  Gill, P. When should we change the definition of the second? *Phil. Trans. Roy. Soc. A* **369**, 4109-4130 (2011).

15  Chou, C. W., Hume, D. B., Rosenband, T. & Wineland, D. J. Optical Clocks and Relativity. *Science* **329**, 1630-1633 (2010).

16  Rosenband, T. *et al.* Frequency Ratio of Al$^+$ and Hg$^+$ Single-Ion Optical Clocks; Metrology at the 17th Decimal Place. *Science* **319**, 1808-1812 (2008).

17  Dehmelt, H. G. Mono-ion oscillator as potential ultimate laser frequency standard. *IEEE Trans. Instrum. Meas.* **IM-31**, 83-87 (1982).

18  Katori, H. Optical lattice clocks and quantum metrology. *Nature Photon.* **5**, 203-210 (2011).

19  Kessler, T. *et al.* A sub-40-mHz-linewidth laser based on a silicon single-crystal optical cavity. *Nature Photon.* **6**, 687-692 (2012).

20  Martin, M. J. *et al.* A Quantum Many-Body Spin System in an Optical Lattice Clock. *Science* **341**, 632-636 (2013).

21  Westergaard, P. G. *et al.* Lattice-Induced Frequency Shifts in Sr Optical Lattice Clocks at the $10^{-17}$ Level. *Phys. Rev. Lett.* **106**, 210801 (2011).

22  Katori, H., Hashiguchi, K., Il'inova, E. Y. & Ovsiannikov, V. D. Magic Wavelength to Make Optical Lattice Clocks Insensitive to Atomic Motion. *Phys. Rev. Lett.* **103**, 153004 (2009).

23  Lodewyck, J., Zawada, M., Lorini, L., Gurov, M. & Lemonde, P. Observation and cancellation of a perturbing dc stark shift in strontium optical lattice clocks. *IEEE Trans. Ultrason. Ferroelectr. Freq. Control* **59**, 411-415 (2012).

24  Hachisu, H. *et al.* Trapping of Neutral Mercury Atoms and Prospects for Optical Lattice Clocks. *Phys. Rev. Lett.* **100**, 053001 (2008).

25  McFerran, J. *et al.* Neutral atom frequency reference in the deep ultraviolet with fractional uncertainty = $5.7 \times 10^{-15}$. *Phys. Rev. Lett.* **108**, 183004 (2012).

26  Derevianko, A., Dzuba, V. A. & Flambaum, V. V. Highly Charged Ions as a Basis of Optical Atomic Clockwork of Exceptional Accuracy. *Phys. Rev. Lett.* **109**, 180801 (2012).

27  Mukaiyama, T., Katori, H., Ido, T., Li, Y. & Kuwata-Gonokami, M. Recoil-limited laser cooling of $^{87}$Sr atoms near the Fermi temperature. *Phys. Rev. Lett.* **90**, 113002 (2003).

28  Takamoto, M. *et al.* Improved frequency measurement of a one-dimensional optical lattice clock with a spin-polarized fermionic $^{87}$Sr isotope. *J. Phys. Soc. Jpn.* **75**, 104302 (2006).

29  Akatsuka, T. *et al.* 30-km-long optical fiber link at 1397 nm for frequency comparison between distant strontium optical lattice clocks. *Jpn. J. Appl. Phys.* **53**, 032801 (2014).

30   Yamaguchi, A. *et al.* Direct Comparison of Distant Optical Lattice Clocks at the 10$^{-16}$ Uncertainty. *Appl. Phys. Exp.* **4**, 082203 (2011).

31   Middelmann, T. *et al.* Tackling the Blackbody Shift in a Strontium Optical Lattice Clock. *IEEE Trans. Instrum. Meas.* **60**, 2550-2557 (2011).

32   Takamoto, M. & Katori, H. Coherence of Spin-Polarized Fermions Interacting with a Clock Laser in a Stark-Shift-Free Optical Lattice. *J. Phys. Soc. Jpn.* **78**, 013301 (2009).